\documentclass[a4paper,12pt]{article}
\usepackage{amsmath}
\usepackage{amsfonts}
\usepackage{authblk}
\usepackage{graphics,epsfig}
\usepackage{float}
\usepackage{xcolor}

\usepackage{color}

\numberwithin{equation}{section}

\usepackage{hyperref}
\hypersetup{
    colorlinks=true,
    linkcolor=blue,
    filecolor=blue,      
    urlcolor=blue,
    citecolor=blue,
}
 
\def\MPl{M_{_{\rm Pl}}^2}

\title{Generation of CMB and Cosmological constant 
via bulk viscosity}
\author[1]{Sanved Kolekar \thanks{sanved.kolekar@cbs.ac.in}}
\author[2]{S. Shankaranarayanan \thanks{shanki@phy.iitb.ac.in}}
\author[1]{S. M. Chitre \thanks{kumarchitre@cbs.ac.in}}
\affil[1]{\it UM-DAE Centre for Excellence in Basic Sciences,
Mumbai 400 098, India}
\affil[2]{\it Department of Physics, Indian Institute of Technology Bombay, Mumbai 400076, India}

\date{\today}
\begin{document}
\maketitle
\begin{abstract}
A simple model of uniformly expanding, homogeneous Universe with a bulk viscosity is studied wherein the inflationary density decays due to viscous dissipation during the expansion phase of the Universe. The model is shown to generate the Cosmic Microwave Background radiation (CMB). We also demonstrate that, at late times, the inflationary density asymptotically approaches a small finite constant value.
\end{abstract}

\section{Introduction}

The uniformity of large-scale structure of the Universe and the Gaussian nature of primordial fluctuations support the existence of an inflationary epoch in the early Universe~\cite{2006-Bassett.etal-RMP,2018-Baumann-Lect}. In most inflation models, at the epoch of exit of inflation, the homogeneous inflaton begins to oscillate about the minimum of its potential. The inflaton decays into other forms of energy, such as that of matter and radiation, eventually giving the particle content of the Standard Model and perhaps even the dark matter. These more familiar forms of matter and radiation must eventually reach thermal equilibrium at temperatures higher than $1~{\rm MeV}$ to recover the successful big-bang nucleosynthesis~\cite{2006-Bassett.etal-RMP}.

Even though inflation has been successful in explaining observations, the nature of the inflaton, however, remains unclear.
The simplest inflationary models are driven by single scalar fields, while high-energy physics models usually involve many scalar fields~(see, for instance, Refs. \cite{2010-Kaiser-PRD}). If several scalar fields are light enough during the inflationary phase, they are referred to as multi-field inflationary models. Such models allow for a richer inflationary behavior, consequently making predictions more complex~\cite{2013-Kaiser-PRL}. One way to circumvent this issue is to use the result of Zimdahl~\cite{1996-Zimdahl-MNRAS}, who demonstrated that two non-interacting perfect fluids (will different cooling rates) could be mapped into an effective single-fluid with a non-vanishing bulk viscosity. Thus, the multi-field models can be mapped to a perfect fluid with a non-zero bulk viscosity. 

Although the early Universe is far from equilibrium and involves dissipative processes, most of the cosmological models ignore the presence of bulk viscosity. It is known that the contribution of bulk viscosity ($\zeta$) to the fluid stress tensor is significant only when there is a measurable unsteady volume-change in the fluid~\cite{1987-Keizer-Book}. In the early Universe, during the inflationary phase, there is a rapid expansion that causes a change in the volume density and consequently leads to the deviation of the normal stresses from its equilibrium value, which may be identified with the generation of non-zero bulk viscosity. To our knowledge, the effect of bulk viscosity in cosmology was investigated by Treciokas and Ellis~\cite{1971-Treciokas.Ellis-CMP} who showed that a fluid with an equation of state of the form:
\begin{equation}
\label{eq:EOS}
\overline{p} = (\gamma - 1) \overline{\rho} - \overline{\zeta} H(t)
\end{equation}
in flat FRW Universe leads to the following scale-factor:
\begin{equation}
a(t)^{\frac{3 \gamma}{2}} = \frac{2}{3 \overline{\zeta}} \left(\exp\left(\frac{3 \overline{\zeta}}{2} t\right) - 1 \right)
\end{equation}
where $H(t) = \dot{a}(t)/a(t)$, $1 < \gamma < 4/3$ and bulk viscosity ($\overline{\zeta}$) is taken to be a constant during the evolution history. Interestingly, the above solution is well-defined for both positive and negative constant values of the bulk viscosity. (See Ref.~\cite{iver} for earlier work relating to negative bulk viscosity.) For other related works on bulk viscous effects during inflation and primordial perturbations, see Refs. \cite{Barrow, Giovannini}.

The bulk viscosity is compatible with the cosmological principle and can also be contemplated to exist at the background level. 
One may regard the bulk viscosity as a measure of the dissipation of energy when the fluid expands. Due to rapid expansion during the early phase of the Universe, the bulk viscosity of the fluids in the Universe could be a function of background density. The presence of bulk viscosity as an arbitrary function of background density has been considered for inflationary solutions and even for dark matter~\cite{1996-Zimhdahl-PRD,2013-Velten.etal-PRD}. In this work, our focus is on the transfer of the inflaton energy due to viscous dissipation. The scenario we consider is the following: The Universe undergoes the standard inflation. After the exit of inflation, the bulk viscosity of the fluid contributes to the transfer of energy from the inflaton to standard model particles. 

In Ref.~\cite{Chitre}, using a perfect fluid with bulk viscosity, it was demonstrated that the CMB is generated as a consequence of the decay of the inflationary energy density at the end of inflation. The model assumed the bulk viscosity to be a specific function of the inflationary density and radiation density. For a closed Universe, the residual inflationary density at the end of the decay was found to asymptotically approach a finite value and identified with the cosmological constant. However, for a  flat Universe $k = 0$, it was found that the inflationary density decays to zero. 

In this work, we revisit the case for a flat $k=0$ uniformly expanding homogeneous Universe with a wide class of functions assumed to describe the bulk viscosity, allowing it to be either positive or negative. We demonstrate the existence of a general family of functions for the bulk viscosity, for which the inflationary density decays on account of the effect of bulk viscosity during the expansion phase of the Universe. Further, we show that the model leads to the generation of CMB, and the inflationary density asymptotically approaches a finite constant value at the end of the decay phase. We also discuss the implications of the energy conditions for such a model. 

In Sec. \eqref{sec:Model}, we discuss the underlying model and derive the fundamental equations using two different approaches. We show that both approaches lead to an identical set of equations. In Sec. \eqref{sol}, we construct a consistent set of solutions that lead to the exchange of energy from inflaton to radiation. In Sec. \eqref{discsection}, we conclude by briefly discussing the importance of these results.

In this work, we use $(+,-,-,-)$ signature for the 4-dimensional space-time. We set $c = 1$ and $\MPl = 1/(8 \pi G)$ is the reduced Planck mass. We denote dot as derivative with respect to cosmic time $t$ and $H(t) = \dot{a}(t)/a(t)$.

\section{Model set-up}
\label{sec:Model}

In this section, we describe a model of the early Universe and discuss two different approaches to study the model. In the first approach, we assume the two components (radiation and inflaton) in the early Universe to interact with each other.  In the second approach, we treat the entire system to be described by a single fluid with a non-zero bulk viscosity. We show that both approaches lead to an identical set of equations of motion. 

\subsection{Approach 1}\label{setup1}

With a view to make an explicit construction, we consider a two-component system with the presence of both inflationary density and radiation density. The two components interact with each other through a transfer of energy-momentum between them via the channel of a bulk viscosity, $\zeta(t)$. We can formally write down a stress energy tensor ${\cal T}^{ab}$ for the system as:
\begin{eqnarray}
{\cal T}^{ab} = {\cal T}^{ab}_{(i)}  + {\cal T}^{ab}_{(r)} = \left( \varrho(t)  + P(t) 
\right) u^a u^b - P(t) g^{ab} \, ,
\label{newtab}
\end{eqnarray}
where 
\begin{equation}
\label{app1:rhoPdef}
\varrho(t) \equiv \rho_i(t) + \rho_r(t);~~ 
P(t) \equiv p_i(t) + p_r(t) \, ,
\end{equation}
and $\varrho(t)$ and $P(t)$ represent the contributions arising from the sum of the individual density $(\rho_i, \rho_r)$ 
and pressure $(p_i, p_r)$. The equations of motion of the system are governed by $\nabla_b{\cal T}^{ab}=0$. 
The first of the Friedman equations, corresponding to the time-time component of the Einstein's equations, is given by
\begin{equation}
H^2(t) + \dfrac{k}{a^2(t)} = \dfrac{\varrho(t)}{3 \MPl}  \, ,
\label{frw1}
\end{equation} 
where $a(t)$ is the scale factor. In this work, we consider the case of $k = 0$. Defining the volume factor $V(t) = (4 \pi/3) a^3(t)$ with (${\dot a}/a =  {\dot V}/(3 V)$), whence Eq.~(\ref{frw1}) takes the form
\begin{equation}
\dfrac{{\dot V}^2(t)}{V^2(t)} = \frac{3}{\MPl}  \varrho(t)
\label{frw1a}
\end{equation} 
The second Friedman equation denoting the space-space component of Einstein's equations takes the form:
\begin{equation}
T \dfrac{dS}{dt} \equiv \dfrac{d}{dt} \left( \varrho V \right) + P \dfrac{dV}{dt} = 0
\label{frw2}
\end{equation}
where $dS/dt$ is the rate of entropy change during expansion of the Universe and $T$ is the instantaneous temperature. Due to the interaction between radiation and inflaton, $\rho_i$  and $\rho_r$ will have additional dynamical equation, namely, $\nabla_b{\cal T}^{ab}_{(i,r)}= Q^a_{(i,r)} $. We then have
{
\begin{eqnarray}
\partial_t \rho_{(i)} + \left( \rho_{(i)}  + p_{(i)}  \right) \frac{{\dot V}}{V} &=& Q_{(i)} \\
\label{qrelation}
\partial_t \rho_{(r)} + \left( \rho_{(r)}  + p_{(r)}  \right) \frac{{\dot V}}{V} &=& Q_{(r)} \nonumber
\end{eqnarray}
with the requirement that $Q^a_{(i)}  + Q^a_{(r)}  = 0$ (see, for instance, Ref.~\cite{malik}) which ensures that even though the individual momentum tensors are not conserved but their total ${\cal T}^{ab} = {\cal T}^{ab}_{(i)}  + {\cal T}^{ab}_{(r)}$ is conserved as required, that is, $\nabla_b{\cal T}^{ab}=0$.} Here, $Q^0_{(i,r)}  = Q_{(i,r)} $ represents the contribution of homogeneous transfer of energy between the two components. The interaction term in general is a function of $H, \rho_{i}, \rho_{r}$ and  $\varrho$~\cite{Interaction}. Assuming that the energy transfer takes place via the bulk viscosity, we can generalize Eq.~\eqref{eq:EOS}, to the following 
\begin{equation}
Q_{(i)} =  \epsilon \tilde{\zeta}(t) H^2(t) 
= \epsilon \zeta(t) \left(\frac{\dot{V}}{V} \right)^2 \, ,
\end{equation}
where the bulk-viscosity ($\zeta(t) = 9 \tilde{\zeta}(t)$) is a function of $\rho_i, \rho_r$, and $\varrho$. Here $\epsilon$ is a non-zero parameter that can be either positive or negative. Note that $Q_{(i)}$ 
is related to the derivative of energy density, hence, for dimensional reasons, the interaction term has to be quadratic in $H(t)$.

For the inflaton field with the equation of state $p_i \simeq - \rho_i$, Eq.~(\ref{qrelation}) reduces to
\begin{equation}
\dfrac{d \rho_i}{dt} =  \epsilon \zeta(t) \dfrac{{\dot V}^2}{V^2} \, .
\label{decay}
\end{equation}
The above equation implies that for the inflationary density to decay to radiation, we must have 
$\epsilon \zeta(t) < 0$. Hence depending on the bulk viscosity being either negative or positive, $\epsilon$ can take positive or negative values. We assume that the signature of $\zeta(t)$ is either positive definite or negative definite and does not flip during the whole evolution of the Universe, thus fixing the value of $\epsilon$ to be negative unity or positive unity respectively. 

{Before we proceed with the other features of bulk-viscosity, we provide the realm of application of our model. Any successful model of inflation must have a graceful exit into the deceleration stage of standard cosmology~\cite{2006-Bassett.etal-RMP,2018-Baumann-Lect,Books}.  Chaotic inflation and slow-roll inflation paradigm avoid the graceful exit problem~\cite{Books}.  In this work, we assume that the inflationary model leads to an exit and does not have a graceful exit problem.

Like in any standard inflationary models, in our case also, we assume that all matter except the scalar field (the inflaton) is redshifted to extremely low densities. Thus, at the exit of inflation, the inflaton must decay into other forms of matter and radiation, eventually giving the particle content of the Standard Model and perhaps dark matter~\cite{Books}. 
In the literature, this process is referred to as reheating~\cite{Books}. In our case, the bulk-viscosity transfers the inflaton energy to radiation. This requirements leads to the above condition ($\epsilon \zeta(t) < 0$) on the bulk-viscosity.}

For the radiation field, choice of $p_r = \rho_r/3$ reduces Eq.~(\ref{qrelation}) to 
\begin{equation}
\dfrac{d\rho_r}{dt} +  \frac{4}{3} \rho_r \frac{{\dot V}}{V}  = - Q_{(i)} =  - \epsilon \zeta(t) \frac{{\dot V}^2}{V^2} = - \frac{d\rho_i}{dt}
\label{rqab}
\end{equation}
Note that the addition of the decay equation with the above still maintains $\nabla_b{\cal T}^{ab}=0$, as required. We will now adopt a different approach by considering an effective fluid with viscosity and demonstrate that this again leads to the above equation. 

\subsection{Approach 2}\label{setup2}

Alternatively, one could assume the Universe 
to be made of a single effective fluid with a bulk viscosity $\zeta(t)$ like in Refs.~\cite{1996-Zimdahl-MNRAS,Chitre,paddy}. The energy-momentum tensor for the effective fluid is taken to be
\begin{equation}
T_{ab} = \left( \rho(t) + p(t) \right) u_a u_b - p g_{ab}  + \zeta(t) (g_{ab} - u_a u_b) \nabla_c u^c
\label{app2:tab}
\end{equation}
where $\rho$ and $p$ represent respectively the density and pressure of the 
effective fluid with a 4-velocity $u^c$. The density $\rho$ is still equal to the sum of the individual densities, i.e. 
$\rho = \rho_i + \rho_r = \varrho$. However, as shown in Ref.~\cite{1996-Zimdahl-MNRAS}, the pressure $p$ is related to the individual pressures and the bulk viscosity, through:
\begin{equation}
p =  P + \zeta(t) \frac{{\dot V}}{V} = p_i + p_r + \zeta(t) \frac{{\dot V}}{V} \, .
    \label{pprelation}
\end{equation}
Note that this is a generalization of the equation of state in Eq.\eqref{eq:EOS}.

The time-time component of Einstein's equation will be independent of the bulk viscosity and lead to Eq.~(\ref{frw1}). The space-space component of the Einstein's equations, however will be sourced by the bulk viscosity term and is given by:
\begin{equation}
T \dfrac{d {\cal S}}{dt} \equiv \dfrac{d}{dt} \left( \rho V \right) + p \dfrac{dV}{dt} = \zeta(t) \dfrac{{\dot V}^2}{V} \, .
\label{app2:frw2}
\end{equation}
Note that the above equation illustrates that the rate of entropy ${\cal S}$ creation of the effective fluid is non-zero, while in the earlier case (\ref{frw1}), the entropy $S$ is a constant. The entropy ${\cal S}$ relates only to the matter domain, as described by the stress-energy tensor in Eq.~(\ref{app2:tab}), which includes the entropy contributed by the effective fluid and the viscous dissipation. However, it does not incorporate the gravitational entropy associated with the microscopic degrees of freedom of the space-time geometry~\cite{graventropy}. We shall return to this point in section \eqref{discsection}.

Further, one can easily check that replacing the effective pressure $p$ with $P$ using the relation in Eq.~(\ref{pprelation}), the second Friedman equation (\ref{app2:frw2}) reduces to Eq.~(\ref{frw2}) in Approach 1. In general, we then have the relation between the two entropies as
\begin{equation}
    T \dfrac{d {\cal S}}{dt} = T \dfrac{dS}{dt} +  \zeta(t) \frac{{\dot V}}{V} \, .
\end{equation}
In this approach we assume, as in Ref.~\cite{Chitre}, that the inflationary density decays are caused on account of forces arising from the bulk viscosity during the expansion phase of the Universe and write
\begin{equation}
\dfrac{d \rho_i}{dt} =  \epsilon \zeta(t) \dfrac{{\dot V}^2}{V^2}
\label{decay2}
\end{equation}
where $\epsilon$ is the same non-zero parameter defined earlier. In this approach, the above decay equation is an additional assumption regarding the dynamics of the inflationary density. The divergence-less-ness  of the stress-energy tensor defined in Eq.~(\ref{app2:tab}) leads to the second Friedman equation (\ref{frw2}). Note that the spatial component of the divergence-free equation vanishes due to homogeneity. 

Using the decay equation (\ref{decay}) for inflationary density, the equation of state relations for $p_i$ and $p_r$ in the second Friedman equation (\ref{app2:frw2}), one obtains the identical equation for radiation density (\ref{rqab}), i. e.,
\begin{equation}
\dfrac{d\rho_r}{dt} +  \frac{4 \, \rho_r}{3}  \frac{{\dot V}}{V}  =  - \frac{d\rho_i}{dt} = - \epsilon \zeta(t) \frac{{\dot V}^2}{V^2}
\label{rqab2}
\end{equation}
Thus, the two approaches are equivalent and lead to identical equations.

The model has one unknown function $\zeta(t)$
and the parameter $\epsilon$ which can be positive or negative. Using the condition that the inflaton energy decays to radiation, we construct a family of functions of $\zeta(t)$ which provide consistent solutions to the evolution equations (\ref{decay}) and (\ref{rqab2}) by demanding that the model satisfies the following two requirements:
\begin{itemize}
\item \textit{Requirement 1:} $\rho_i$ decays from a maximum value $\rho^{max}_i$ to  a small residual non-negative value denoted by $\rho_{\lambda}$  \label{req1}
\item \textit{Requirement 2:} $\rho_r$ is created as a result of the decay of $\rho_i$, that is, starting from a null value in density, $\rho_r$ reaches a max value $\rho_r^{max}$ and eventually decays to the observed present value of the CMB, $\rho_r^{CMB}$ on account of the expansion of the Universe. \label{req2}
\end{itemize}

These requirements are consistent with the general perception that at the end of the inflationary era, i.e., the dominant contributions to the energy density of the Universe comes from the inflaton with the equation of state $p_i \simeq - \rho_i$. The radiation density $\rho_r$ is negligible compared to $\rho_i$ at this stage in the evolution of the Universe. For simplicity, we shall assume that the radiation density to be zero initially.

One should note from the explicit construction described above that we do not attempt to replace the standard inflationary scenario with that from the bulk viscosity. We have used the quasi de Sitter equation of state for the inflaton, i. e., $p_i \simeq - \rho_i$ to arrive at the form of Eq.~(\ref{decay}). The scenario we consider is the following: The Universe undergoes the standard inflation, and after the exit of inflation, the bulk viscosity of the fluid contributes to the transfer of energy from the inflaton to standard model particles. This is discussed explicitly in Sec. 3.1 for two different bulk viscosity functions. 

\section{General solution}
\label{sol}

Given the essential details about the model, in this section, we construct a family of consistent solutions that satisfy the above two requirements. Using Eqs.~(\ref{frw1}) and (\ref{rqab}), we get
\begin{eqnarray}
\dfrac{d\varrho}{dt} &=&  - \dfrac{4}{\sqrt{3 \MPl}} \left(\rho_r \right) \sqrt{\omega \varrho} \, . 
\label{totaleqn}
\end{eqnarray}
where $\omega = 24 \pi G$. Dividing the above equation by the decay equation for the inflationary density Eq.~(\ref{decay}), we get,
\begin{equation}
\dfrac{d \varrho}{d \rho_i} = - \dfrac{4}{3 \epsilon \zeta(t) } \left(\dfrac{\varrho - \rho_i}{\sqrt{ \varrho}} \right) \, .
\label{densityspace}
\end{equation}
For a given bulk viscosity function $\zeta(\rho_i(t),\varrho(t)) $, one can then find a solution corresponding to the trajectory in the 2-dimensional density space $\{ \rho_i, \varrho \}$. In Ref.~\cite{Chitre}, one such family of solutions was found for a particular choice of $\zeta\propto (\varrho - \rho_i)\sqrt{\rho_i/\varrho}$ for closed FRW space-time. 

Here, we do not make any assumption about the form of $\zeta$, and \emph{only} about the asymptotic properties of $\zeta$. We substitute the decay equation for the inflationary density Eq.~(\ref{decay}) in Eq.~(\ref{totaleqn}), we get:
\begin{equation}
\dfrac{d\rho_r}{dt} =  - \dfrac{4 \rho_r}{\sqrt{3 \MPl}} \sqrt{ \rho_i + \rho_r} - \dfrac{d\rho_i}{dt}
\label{requation}
\end{equation}
We would like to draw the {readers attention} to an important feature from the above equation: The above equation is sufficient to show that if the first requirement is imposed, then the second requirement is automatically satisfied for a range of values of the parameter $\epsilon$ which we shall determine. As per the first requirement, let us start with the minimal assumption that the inflationary density is a monotonically decreasing smooth differentiable function, which asymptotes to a small finite value  $\rho_{\lambda}$ as $t \to \infty$. $d\rho_i/dt$ is also monotonically decreasing smooth differentiable function and is negative; hence, it asymptotes to zero as $t \to \infty$. 

Let us look at the two terms in the RHS of the above evolution separately: The first term in the right-hand side is always negative in magnitude, and since $\rho_r^{\rm initial} =0$; initially, it is zero. This term is negligible in the initial evolution of the radiation density. The second term, which is the rate of change of inflationary density ($\rho_i$), is always positive as it decays (negative). Initially, the second term dominates. It is a monotonically decreasing smooth differentiable function which asymptotes to zero as $t$ goes to infinity. The combined effect of both of the terms is the following: $\rho_r$ initially increases because of the second term, then reaches a maximum value when both the terms become equal and opposite in magnitude and then gradually decreases when the first term starts to dominate over the second term. 

Thus, the radiation density is generated at the expense of the decay of the inflaton. The radiation density then falls off due to the expansion of the Universe. At late times, the fall-off can be determined by ignoring the small decay term and by replacing $\rho_i$ by $ \rho_{\lambda}$ in the first term. For a small $\rho_{\lambda} \ll 0$, the radiation density falls off as $\rho_r \propto t^{2}$ as to be expected.

We have thus qualitatively shown that the existence of first requirement  automatically implies the satisfaction of the second requirement. The constraint on the parameter $\epsilon$ determines the overall sign of $\zeta(t)$. For $\epsilon = -1$, the bulk viscosity is positive, while for $\epsilon = 1$, bulk viscosity is negative. For the first requirement to hold,  the bulk viscosity needs to satisfy 
\begin{equation}
\zeta(t) = \dfrac{-\epsilon}{|\epsilon|^2 \omega (\rho_r + \rho_i)} F(g^{-1}(\rho_i))
\label{zetadef}
\end{equation}
where $F(\sigma)$ is a smooth differentiable negative function with the constraint that it approaches zero from below faster than $-1/\sigma$ when $\sigma$ goes to infinity. The function $g(\sigma)$ is determined through $F(\sigma)$ as $g(\sigma)  = \int F(\sigma) d\sigma $. It is then straightforward to check that $\rho_i$ in Eq.~(\ref{decay}) is simply $\rho_i(t)  = \int F(t) dt = g(t)$ with appropriate boundary conditions so as to satisfy the first requirement. Thus we have obtained the general solution for the bulk viscosity in Eq.(\ref{zetadef}) which satisfies both the requirements 1 and 2. 

\subsection{Two examples for bulk viscosity}

\begin{figure}[!htb]
\centering
\includegraphics[width=0.7 \linewidth]{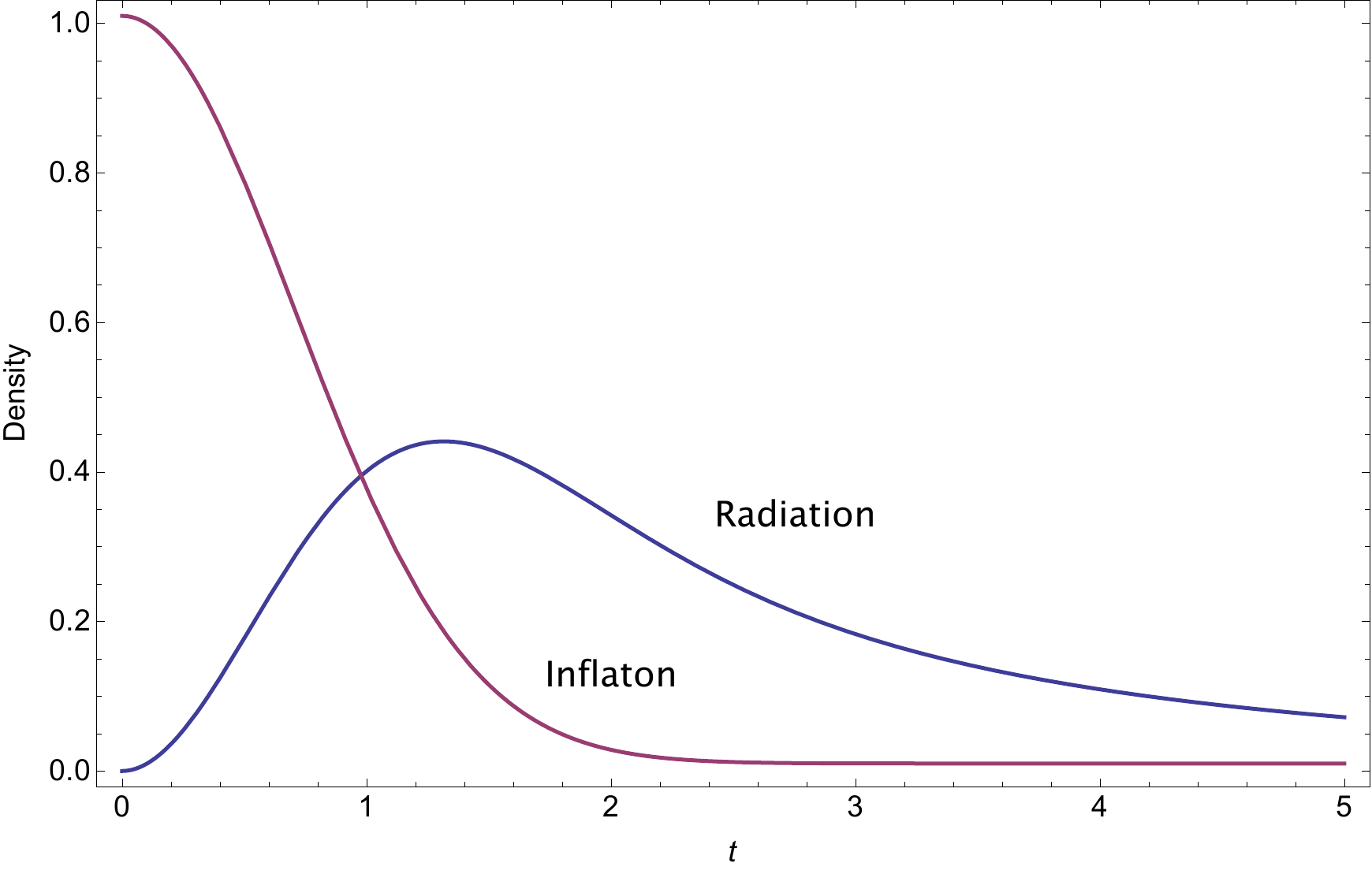}
\caption{For a choice of $\zeta(t)$ given in Eq.(\ref{zeta2}), the densities $\rho_i$ and $\rho_r$ are plotted as a function of time $t$. Here $\rho_{\lambda} = 0.01$
\label{fig:diagram2}}
\end{figure}

Below, we consider two examples for the form of bulk viscosity based on the general solution in Eq.(\ref{zetadef}). One should note that these forms are for illustrative purposes only. 
\begin{enumerate}
    \item Consider the following form of $F$, namely,
    $$F(\sigma) = - \rho_{\lambda} \frac{\exp(-t)}{(c - \exp(-t))^2} \, ,$$
    where $\rho_{\lambda}$ and $c$ are constants with $c > 1$ and $t$ is dimensionless rescaled time. Substituting the above form of $F$ in Eq.~\eqref{zetadef}, we get,
\begin{equation} 
\zeta(t) = \dfrac{-\epsilon}{|\epsilon|^2 \omega (\rho_r + \rho_i)} \left[ \dfrac{\rho_i}{\rho_{\lambda}} -  \dfrac{c \rho_i^2}{\rho_{\lambda}^2}\right] \, .
\label{zeta1}
\end{equation}
Substituting the above expression in Eq.~\eqref{decay}, we get,
$$
\rho_i = \rho_{\lambda} \frac{\exp(t)}{c \exp(t) - 1} \, .
$$
Substituting the above expression in \eqref{requation}, we obtain the time dependence of the radiation density. Figure~\ref{fig:diagram1} contains the plot of radiation and 
inflaton densities as a function of $t$. From the figure, it is clear that initially, radiation was negligible, and due to bulk viscosity, the inflaton energy gets transferred to radiation energy. At an epoch, say $t = t_*$, the radiation energy density is maximum when the inflaton energy has reached a minimum value and continues to remain in that value during the expansion. 

\item Consider the second form of $F$, namely,
$$F(\sigma) = -2t \exp(-t^2)$$
Substituting the above form of $F$ in Eq.~\eqref{zetadef}, we get,

\begin{figure}[!htb]
\centering
\includegraphics[width=0.7 \linewidth]{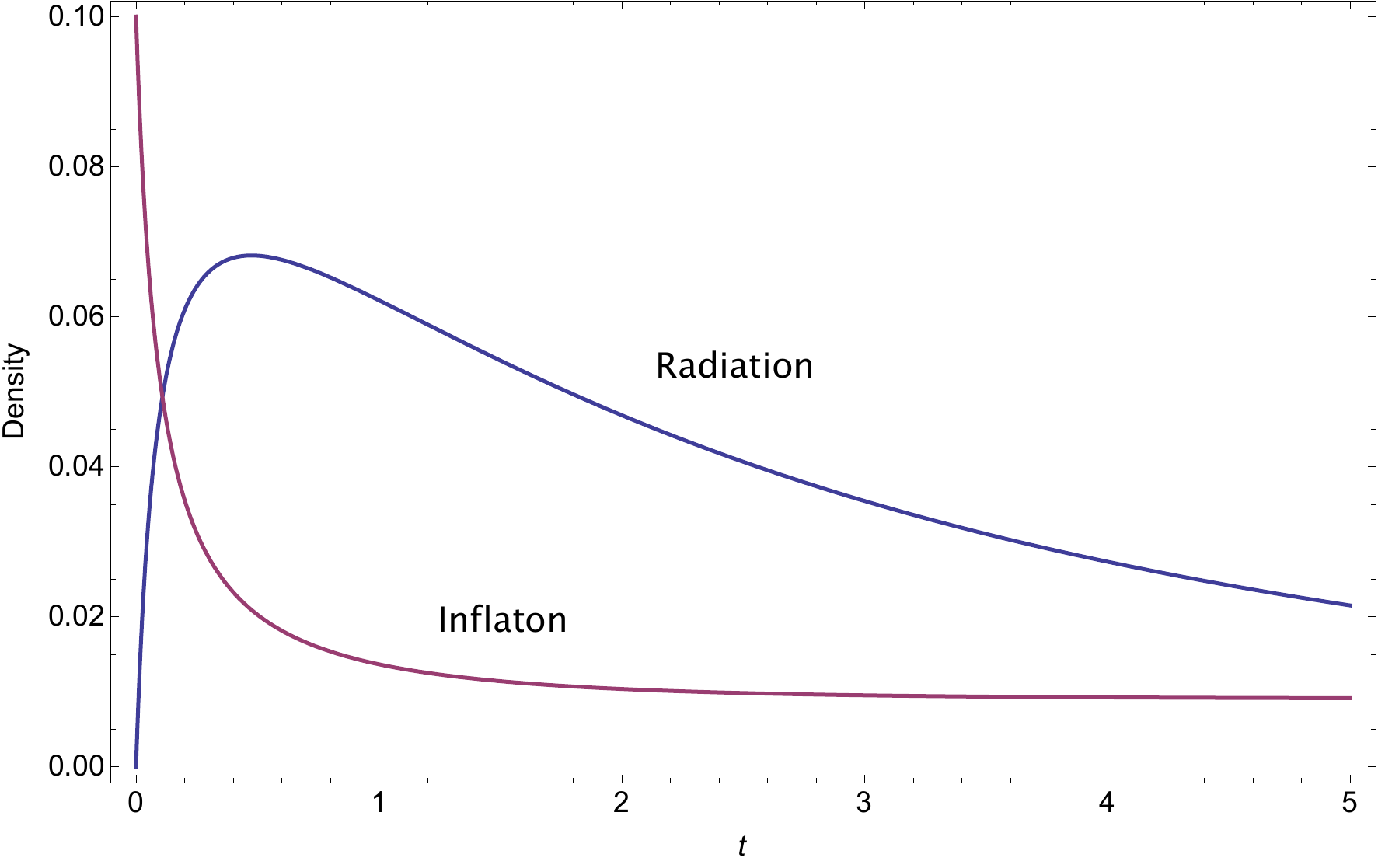}
\caption{For a choice of $\zeta(t)$ given in Eq.(\ref{zeta1}), the densities $\rho_i$ and $\rho_r$ are plotted as a function of time $t$. Here $\rho_{\lambda} = 0.01$ and $c = 1.2$
\label{fig:diagram1}}
\end{figure}

\begin{equation}
\zeta(t) = \dfrac{2 \epsilon (\rho_i - \rho_\lambda) }{|\epsilon|^2 \omega (\rho_r + \rho_i)} \sqrt{- \log{\left( \rho_i - \rho_{\lambda} \right)}}
\label{zeta2}
\end{equation}
Repeating the earlier analysis, we get,
$$\rho_i =  \exp{(-t^2)} + \rho_{\lambda}$$
In other words, the inflaton energy density decays exponentially with a residual value at late time. 
Figure \ref{fig:diagram2} contains the energy density
of the inflaton and radiation as a function of $t$. 
Here again, we see that the radiation density peaks at an epoch ($t = t_*$).
\end{enumerate}

Thus, the radiation density in the Universe is generated at the expense of decay of the inflaton field, leaving a residual constant energy density, which is identified with the cosmological constant. Our analysis is consistent with Hawking's 
result~\cite{1970-Hawking-CMP} where he  showed that in classical General Relativity if the space-time 
is non-empty at an initial time, it will remain non-empty at all times, provided it satisfies the stress-tensor 
conservation.

\subsection{Energy conditions}

Given the generic form of the bulk viscosity function (\ref{zetadef}), it is necessary to investigate the implications of the energy conditions on the feasibility of such a choice. To proceed further, we re-arrange the pressure and $\zeta(t)$ terms in  Eq.~(\ref{app2:tab}) in the form of an ideal fluid tensor:
 \begin{equation}
T_{ab} = \rho(t)  u_a u_b -  (g_{ab} - u_a u_b) \left[ p(t) - \zeta(t) \nabla_c u^c \right]
\label{tab2}
\end{equation}
Thus the effect of bulk viscosity in an expanding Universe is to stifle or build-up pressure depending on the sign of $\zeta(t)$. The weak energy condition implies that 
$$ 
\rho(t) \geq 0;~{\rm and}~\rho(t) + p(t) - \zeta(t)\frac{\dot V}{V}  = \rho(t) + P(t) \geq 0
$$ 
The first condition is satisfied by definition since, in the present case, all the densities considered are non-negative. The second condition is also satisfied when one notes that for the FRW background, the effective pressure $p$ can be written as 
\[P(t) = p(t) - \zeta(t) \frac{{\dot V}}{V} \, . \] 
Since $\rho + P = (4/3) \rho_r$ is always non-negative, the weak energy condition is trivially satisfied for both positive and negative signs of the bulk viscosity $\zeta(t)$. 

\section{Discussion}\label{discsection}

We have established that radiation density can be generated at the expense of the decay of the 
inflaton via the bulk viscosity in an expanding Universe at the end of inflation. The inflaton's potential energy drives the dynamics of inflation, and the bulk viscosity does not play a role during the inflationary epoch. The scenario we have considered is the following: The Universe undergoes the standard inflation, and after the exit of inflation, the bulk viscosity of the fluid contributes to the transfer of energy from the inflaton to standard model particles.   A general form of the bulk viscosity was determined in Eq.~(\ref{zetadef}) depending on the densities ($\rho_i, \rho_r$), 
and constrained only by its asymptotic fall-off properties. Remarkably, the bulk viscosity can be positive or negative, with a specific value of its coupling with the decay equation such that weak energy condition is satisfied at all times. 

The rate of entropy generation or its depletion in Eq.~(\ref{app2:frw2}) depends on the sign of bulk viscosity. For a definite positive signature for $\zeta(t)$, the effective fluid's entropy generation would increase. In contrast, it would seem that the total entropy decreases for a 
negative choice of $\zeta(t)$. However, as stressed earlier, this entropy only considers 
the entropy content in the matter domain, as described by the stress-energy tensor defined in 
Eq.~(\ref{newtab}). It is widely believed that the space-time geometry also contributes to the entropy due to the quantum gravitational degrees of freedom describing the geometry as a 
long-wavelength limit of the underlying, yet to be formulated a complete theory of quantum gravity~\cite{graventropy, paddyaseem}. For example, a black hole is known to have an 
entropy associated with itself, which equals $1/4$th value of the area of its horizon measured in 
Planck area units, while obeying the Generalised second law of thermodynamics requiring the total entropy of the matter domain and the black hole entropy to increase always in an irreversible process \cite{bh}. One should then expect a similar inequality stating that the entropy $S_{m}$ 
in the matter domain and the entropy $S_{g}$ in the background gravitational domain to obey 
$dS_m/dt + dS_g/dt \geq 0$ during expansion of the Universe. Thus, $S_g$ could grow at the expense of a decreasing $S_m$ while ensuring their total rate of change to be always 
non-negative~\cite{2010-Egan.Lineweaver-ApJ}.

In our analysis, the bulk viscosity is the primary catalyst for the exchange of energy between the inflaton and the radiation field in the early Universe. Although our model presumes the two fluids to interact only through the background space-time, it has been implicitly assumed that they are indirectly coupled through the functional dependence of bulk viscosity on the densities. 
The microscopic description starting from a Lagrangian for such a system of fluids corresponding 
to the stress-energy tensor in Eq.~(\ref{newtab}) is a highly non-trivial task given the fact that the 
Lagrangian description of the familiar Navier-Stokes equation remains elusive~\cite{san}. 
The underlying model outlined in this work is primarily classical and is in a spirit similar to that of 
Bateman-Feshbach-Tikochinsky (BFT) oscillator theory~\cite{DHO}. BFT oscillator is made up of
two oscillators where the energy is transferred from one to the other with the Hamiltonian corresponding to the BFT oscillator itself being conserved. 

The bulk viscosity in Eq.(\ref{zetadef}) is a function of the energy densities of both the inflationary and radiation fields. If we were to attempt to define the Lagrangian dynamics for the corresponding non-conservative dissipative system, which involves an energy transfer between the two components of the system, then there would be an interaction term between the two fields in the Lagrangian. One would then expect a similar interaction energy term in the $T_{uu}$ component of the stress-energy tensor. A Lagrangian based field-theoretic description for interacting fluids shown in Refs. \cite{Gonzalez,Johnson:2020gzn} can describe the approach mentioned in section \ref{setup1}.
These aspects need to be investigated, and we hope to address it in future communication.

The small residual asymptote, $\rho_\lambda $ of the inflationary density at late times, is a consequence of our choice of the bulk viscosity function in Eq.~(\ref{zetadef}). It arises as 
one of the parameters in the definition of the function $F(\sigma)$ appearing in Eq.(\ref{zetadef}). 
One could, in principle, try to put constraints on the parameter $\rho_\lambda$ by demanding the value of radiation density in the solution to match that of the CMB at present epoch and then fit the curves for a suitable value of $\rho_\lambda$. However, one notices there are other parameters such as the initial inflationary density at $t =0$ and the slope of $F(\sigma)$ at early times to which the radiation curve at late times is sensitive to and thus setting satisfactory bounds on $\rho_\lambda$ may not be feasible. Our analysis provides a plausible way of understanding the smallness of the cosmological constant! 

The current analysis can be extended to include matter and other components that we hope to report shortly. 

\section*{Acknowledgements}
SMC is grateful to the late Donald Lynden-bell for resurrecting his interest in the role of bulk viscosity in the context of cosmological models. The authors thank T. Padmanabhan for useful comments on an earlier version of the manuscript. SK thanks the Department of Science and Technology India for financial support under the Inspire Faculty grant. The ISRO-Respond grant partially supports the work of SS.

\end{document}